\documentclass[sigconf]{aamas}
\usepackage{balance} % for balancing columns on the final page

\usepackage{amsmath}
\usepackage{graphicx}
\usepackage{booktabs}
\usepackage{caption}
\usepackage{array}
\usepackage{amsfonts}

\setcopyright{none}

\acmConference[AAMAS '21]{Proc.\@ of the 20th International Conference on Autonomous Agents and Multiagent Systems (AAMAS 2021)}{May 3--7, 2021}{Online}{U.~Endriss, A.~Now\'{e}, F.~Dignum, A.~Lomuscio (eds.)}
\copyrightyear{2021}
\acmYear{2021}
\acmPrice{}
\acmISBN{}

\title[Cooperation and Reputation Dynamics with Reinforcement Learning]{Cooperation and Reputation Dynamics \\ with Reinforcement Learning}

\author{Nicolas Anastassacos}
\affiliation{
  \institution{University College London}}
    \affiliation{
  \institution{The Alan Turing Institute}}
\email{nanastas@cs.ucl.ac.uk	}

\author{Julian Garc\'ia}
\affiliation{
  \institution{Monash University}}
\email{julian.garcia@monash.edu}

\author{Stephen Hailes}
\affiliation{
  \institution{University College London}}
\email{s.hailes@ucl.ac.uk}

\author{Mirco Musolesi}
\affiliation{
  \institution{University College London}}
      \affiliation{
  \institution{The Alan Turing Institute}}
    \affiliation{
  \institution{University of Bologna}}
\email{m.musolesi@ucl.ac.uk}

\begin{abstract}
Creating incentives for cooperation is a challenge in natural and artificial systems.  
One potential answer is reputation, whereby agents trade the immediate cost of cooperation for the future benefits of having a good reputation. 
Game theoretical models have shown that specific social norms can make cooperation stable, but how agents  can independently learn to establish effective reputation mechanisms on their own is less understood. 
We use a simple model of reinforcement learning to show that reputation mechanisms generate two coordination problems: agents need to learn how to coordinate on the meaning of existing reputations and collectively agree on a social norm to assign reputations to others based on their behavior. 
These coordination problems exhibit multiple equilibria, some of which effectively establish cooperation. 
When we train agents with a standard Q-learning algorithm in an environment with the presence of reputation mechanisms, convergence to undesirable equilibria is widespread. 
We propose two mechanisms to alleviate this: \emph{(i)} seeding a proportion of the system with fixed agents that steer others towards good equilibria; and \emph{(ii)}, intrinsic rewards based on the idea of introspection, i.e., augmenting agents' rewards by an amount proportionate to the performance of their own strategy against themselves. 
A combination of these simple mechanisms is successful in stabilizing cooperation, even in a fully decentralized version of the problem where agents learn to use and assign reputations simultaneously. 
We show how our results relate to the literature in Evolutionary Game Theory, and  discuss implications for artificial, human and hybrid systems, where reputations can be used as a way to establish trust and cooperation.
\end{abstract}


\DeclareMathOperator*{\argmax}{arg\,max}

\newenvironment{psmallmatrix}
  {\bigg(\begin{smallmatrix}}
  {\end{smallmatrix}\bigg)}
\begin{document}

%%% The following commands remove the headers in your paper. For final 
%%% papers, these will be inserted during the pagination process.

\pagestyle{fancy}
\fancyhead{}

%%% The next command prints the information defined in the preamble.

\maketitle 

%%%%%%%%%%%%%%%%%%%%%%%%%%%%%%%%%%%%%%%%%%%%%%%%%%%%%%%%%%%%%%%%%%%%%%%%

\section{Introduction}

%\item The problem of cooperation
Cooperation is important in natural and artificial systems \cite{axelrod:1997}. 
It allows for agents with individual goals to reach beneficial group outcomes, even when group and individual incentives are not perfectly aligned \cite{sigmund:2016}.
If cooperation is costly but the benefits of cooperation can be enjoyed by all agents, the temptation to pay no cost is a dominant strategy and cooperation is hard to establish and maintain, unless a specific mechanism is in place to foster cooperation \cite{nowak:2006a}.

One popular mechanism is direct reciprocity, which allows for agents to meet repeatedly \cite{vanveelen:2012, garcia:2018}, thereby creating incentives to punish past defections, making cooperation viable via reciprocal strategies like Tit-for-Tat \cite{axelrod:1981}. 
When agents are anonymous or cannot interact repeatedly, they can use a reputation mechanism to condition their cooperative actions, e.g., cooperating only with those that have a good reputation.
This is known as indirect reciprocity \cite{santos:2016, santos:2018}, and allows agents to trade the immediate cost of cooperation for the future benefits of keeping their reputation \cite{mailath:2006}. 

Reputation systems are common in computing, including multiagent systems \cite{vogiatzis:2010,granatyr:2015}, but their application is not always straightforward \cite{jurca:2003, hoffman:2009, santos:2018a}. Models of indirect reciprocity can be used as tools to understand reputation-based systems \cite{ohtsuki:2006, santos:2018, xu:2019, krellner:2020}. These models generally provide a mathematical description of the incentives, coupled with a dynamic account of how groups of agents respond to these incentives in simple but illustrative scenarios. The dynamic features of these models are crucial, given that static models might be insufficient in the presence of multiple equilibria~\cite{mailath:2006}.

%rephrase
The framework of indirect reciprocity typically relies on the toolset of Evolutionary Game Theory (EGT).
In EGT, agents do not solve for equilibrium, but copy other agents that are successful in a dynamic process inspired by evolution~\cite{sigmund:2016}. Agents that perform well in a population are more likely to pass on their strategies or policies to subsequent generations.
Despite their simplicity, these models have been successful in predicting human behavior \cite{hoffman:2015} and to contribute to our understanding of complex dynamics associated with learning algorithms in multiagent settings~\cite{tuyls:2003, tuyls:2005, sen:2007, hennes:2020}.

In models of reputation dynamics, agents learn how to take into consideration the reputations of others from rewards derived from a series of interactions with randomly chosen partners \cite{nowak:2005}. A strategy determines whether a co-player with a particular reputation is worth the cost of cooperation. A coupled dynamic process emerges from the changing reputations in the system. 
Social norms determine how agents judge interactions, updating the reputations of other agents after each encounter. Therefore, a norm assigns a reputation value given a combination of factors. Game theoretical models show that norms that reward justified defection, as well as cooperation with reputable agents, are particularly good at maintaining cooperation \cite{ohtsuki:2006, nowak:2005, santos:2016}.

Crucially, EGT models assume that the set of strategies are predefined, and explored by agents in a  random fashion. 
Instead, we use Reinforcement Learning (RL) to model the process whereby agents discover strategies from simpler states and actions.
Agents explore and learn strategies (i.e., policies) responding directly to the rewards obtained during game play.

Our purpose is twofold: 
First, we ask how predictions of models of reputation dynamics change, when the learning process is driven by individual experiences like in RL, instead of social learning in EGT models. 
Second, we provide a simple environment where we test whether RL algorithms reliably learn the equilibria that lead to sustained cooperation on the basis of public reputations.

Studies of RL and cooperation abound (see, for example \cite{banerjee2003mutual, foerster:2018, letcher:2018, hughes:2018, jaques:2019, eccles:2019, anastassacos:2020}, but the interaction between cooperation and reputation in this context has not been explored yet.
The existing EGT literature on reputation can be insightful in analyzing and proposing solutions to this particular problem \cite{tuyls:2005}. This is the first work that bridges RL and EGT literature in the area of cooperation and reputation dynamics. In particular, we find that in presence of reputation mechanisms, agents need to solve two coordination problems: learning how to coordinate actions on the basis of existing reputation indicators; and collectively agreeing on a social norm to assign reputations to others based on their behavior. 
These coordination problems exhibit multiple equilibria, some of which effectively establish cooperation.
Agents driven by standard RL algorithms will generally fail to coordinate, converging to non-efficient outcomes. 
This can be resolved by seeding a proportion of the system with fixed agents that steer others towards good equilibria or providing intrinsic rewards based on the idea of introspection, i.e., augmenting agents' rewards by an amount proportionate to the performance of their own strategy against themselves. 

More generally, our results can also inform the work of AI researchers concerned with designing agents that can cooperate amongst themselves \cite{eccles:2019}, and with humans \cite{santos:2019a, correia:2019a}.
The problem of cooperation is also present in scenarios where artificial agents respond to individual rewards \cite{peysakhovich:2018, rahwan:2020, foerster:2018, anastassacos:2020}.
Because individual incentives are often not sufficient, intrinsic rewards \cite{hughes:2018, jaques:2019, eccles:2019} or more complex agent architectures where the choice of the next action is based on prediction models of the behavior of the other agents \cite{foerster:2018, letcher:2018}. 
We show that, fundamentally, a reputation mechanism is not sufficient to steer agents towards cooperation, but this can be combined with intrinsic rewards to achieve cooperation.

The rest of this paper is organized as follows: Section \ref{preliminaries} describes the problem of cooperation in presence of reputation mechanisms and the existing results based on stability solution concepts. 
Section \ref{easyproblem} shows that naive Q-learning agents converge to undesirable equilibria.
Section \ref{fixes} discusses how to steer the system towards desirable outcomes while Section \ref{hardproblem} deals with how agents can learn to also assign reputations in a completely decentralized fashion. 
In Section \ref{discussion}, we discuss how our results relate to the existing literature in game theory, as well as implications for artificial, human and hybrid systems, where reputation can be used as a way to establish trust and cooperation.

\section{Preliminaries \label{preliminaries}}

\subsection{Prisoner's Dilemma}

% A wealth of game theoretical literature, both in AI and GT
Our investigation is based on the classic Prisoner's Dilemma (PD) game \cite{rapoport:1965}.
Agents can cooperate, paying a cost $c$ to help their opponent by an amount $b$; or defect, bypassing the cost and potentially reaping the benefit bestowed from cooperative co-players. 
The corresponding payoff matrix is then: $\begin{psmallmatrix}0 & b\\-c & b-c\end{psmallmatrix}$, with the first action being defect, and the second action cooperate. In the following section we denote defect as action 0 and cooperate as action 1.
With $b > c >0$, selfish agents will play defect; an outcome that is not Pareto optimal given that everyone would better off with mutual cooperation. 
We will set $c=1$, and vary $b$ to adjust the benefit to cost ratio of cooperation.
The amount of cooperation that emerges will be dependent on the benefit to cost ratio of the game $b/c$.

\subsection{Reputation Mechanisms} \label{thegame}
% \item Description of the environment
We investigate a setup typical of the EGT literature \cite{ohtsuki:2006, nowak:2005}.
We consider $N$ agents that are randomly matched with another agent, each round, to play a game. The game is repeated for a number of rounds, with agents being rematched in every round.
An episode lasts for a pre-determined number of rounds $M$.

% \item Prisoner's Dilemma
In each round agents play the PD game described above.
Agents decide whether to cooperate or not based on the reputations of their co-players. 
In our simplest model, reputations can be $0$ or $1$.
For the sake of clarity we will sometimes refer to $1$ as \emph{Good}, and $0$ as \emph{Bad}; however, no meaning is ascribed to reputation values a priori. 
Following \cite{ohtsuki:2004}, an agent decides whether they cooperate or not based on their own reputation and the reputation of the opponent. The reputation assignment is based on a social norm as discussed in Section \ref{socialnorms}.  Thus, we can encode how they react to the reputations of other agents with $4$ bits as shown in Table \ref{tab:action_rules}.

\begin{table}[!h]
%\captionsetup{font=small}
\center
\caption{Action rules are encoded as bitstrings of size 4. Each bit encodes the action of the focal agent, with  $\{D=0, C=1\}$. This action is a function of her reputation and the reputation of the opponent. There are 16 possible action rules.}
\begin{tabular}{@{}|l|l|l|l|l|@{}}
\hline
If the focal reputation is:        & $0$                        & $0$                        & $1$                     & $1$                     \\
and the opponent reputation is: & $0$                             & $1$                             & $0$                             & $1$     \\
\hline
The focal action is given by:                     & Bit 3 & Bit 2 & Bit 1 & Bit 0 \\ 
\hline
%\hline  
 %Rule 5: & D (0) & C (1) & D (0) &  C (1)\\
%\hline                       
\end{tabular}

\label{tab:action_rules}
\end{table}

For example, action rule $5$ is as follows: Bit $3$ is $0$, therefore an agent using this rule will defect if their reputation is $0$ and the opponent's reputation is $0$; Bit $2$ is $1$, thus the agent will cooperate if their reputation is $0$ and the opponent's is $1$; Defect if their own reputation is $1$ and the other's is $0$, as given by Bit $1$, and cooperate when their own reputation and the co-player's reputation is $1$ as given by Bit $0$. The resulting bitwise representation is $\mathbf{0101_2}$. In a similar fashion, action rule $0$ is always defect; action rule $15$ is always cooperate, and so on.
As a result, we have $16$ possible action rules in total. 

% \item Norms

\subsection{Social Norms and Reputation Assignment}
\label{socialnorms}
Now that we have described the basic setup of the game and the concept of action rule, we can define how reputations are assigned to agents after observing social interactions.

A social norm~\cite{wooldridge:2009} is used to determine how agents assign reputations to others \cite{ohtsuki:2004}.
A social norm is a function that translates how the actions of the parties involved in an interaction, translate into their future reputations \cite{ohtsuki:2006}. 
An observer, sometimes centralized, changes the reputations of both parties, following the social norm, after each interaction.

\begin{table}[!h]
%\captionsetup{font=small}
\center
\caption{Bitwise interpretation of norms. Each bit encodes the new focal reputation given her action towards an opponent with a particular reputation. 
 For example, norm $1 = {0001}_2$ assigns a reputation 1 only to agents that cooperate when facing an agent with reputation 1.}
\begin{tabular}{@{}|l|l|l|l|l|@{}}
\hline
If the focal action is:        & $D$                        & $D$                        & $C$                     & $C$                     \\
The opponent's reputation is: & $0$                             & $1$                             & $0$                             & $1$     \\
\hline
New focal reputation is given by :                        & Bit 3 & Bit 2 & Bit 1 & Bit 0 \\ 
\hline
\end{tabular}

\label{tab:norm_bitwise}
\end{table}
%
% \item What are social norms 
% \item 

Following \cite{ohtsuki:2004}, the new reputation for an agent depends on her action, and the reputation of the co-player. 
We can encode a social norm as four bits, for a total of $2^4$ norms (see Table~\ref{tab:norm_bitwise}). 
These are known as second-order norms \cite{nowak:2005}, and can be extended to depend as well on the reputation of the focal agent (third order), or even on the basis of previous interactions \cite{santos:2018}.  Table~\ref{tab:example_norms} gives some social norm examples. 
We consider a small reputation assignment error $\chi$, which occasionally flips a reputation from the original intention on assignment.
This small error allows for the reputation dynamics to be stationary, in the sense that in the long run the effect of initial reputations vanish.

\begin{table*}
\centering
\caption{Examples of social norms.}
\begin{tabular}{@{} c c p{9cm} @{}}
\toprule
%\rowcolor[HTML]{EFEFEF} 
\textbf{Norm} & \textbf{Binary representation} & \textbf{Meaning}                                                                     \\ \midrule
$0^{*}$    & $\mathbf{0000_2}$      & Actions and reputations play no role - always assigns ``bad'' reputation.                                                   \\ \midrule
$3$   & $\mathbf{0011_2}$      & Cooperating with others is always ``good'' and defecting is always ``bad''.                     \\ \midrule
$9^{*}$ & $\mathbf{1001_2}$ & Someone who cooperates with others that are ``good'' and defects to others that are ``bad'' is good.
	\\ \midrule
11   & $\mathbf{1011_2}$    & Someone is ``bad'' only if they refuse to cooperate with a good individual.  \\ \bottomrule
\end{tabular}
\label{tab:example_norms}
%\end{table}
\end{table*}

We consider two scenarios of reputation assignment that reflect different levels of centralization.
In a semi-centralized system all agents used a fixed exogenous norm (top-down reputation). In this case, the system's state is given by $N$ action rules $p_i$, where each action rule is in $\{0.. 15\}$. Instead, in a fully decentralized system each agent can use a different norm (bottom-up reputation) \cite{xu:2019}. 
In this case, the system's state is given by $N$ tuples $(p_i, d_i)$, where each action rule $p_i$ is in $\{0.. 15\}$ and each social norm $d_i$ is in $\{0.. 15\}$.
Section \ref{easyproblem} and \ref{fixes} will deal with top-down dynamics, and Section \ref{hardproblem} will discuss the bottom-up case.

\subsection{EGT Stability Predictions \label{stability}}

In EGT, pairs $(p, d)$ have been analyzed for stability \cite{ohtsuki:2006}. 
Technically speaking, a norm $d$ yields a dynamical system that determines the proportion of ``Good'' individuals in the population in the long run.
This proportion is then used to compute the payoffs of specific action rules.
A norm $d$ stabilizes cooperation, if a monomorphous population using action rule $p$ and norm $d$ can resist invasions by any mutant action rule $p'$.
Social norms that reward justified defection, as well as cooperation with reputable agents, are particularly good at maintaining cooperation.
These results have been extended to more realistic stochastic systems with finite populations \cite{santos:2016}.

Stability predictions show that social norms that are most successful at  cooperation share two characteristics: (i) cooperation with ``good'' agents begets a good reputation, and (ii) defection against ``bad'' agents begets a good reputation \cite{ohtsuki:2006}. 
A particularly salient norm is known as ``Stern Judging" \cite{pacheco:2006}.
The binary representation of this is norm $9$ given in Table \ref{tab:example_norms}.
More specifically, the combination of action rule $5$ and norm $9$ cannot be invaded once established (see Tables \ref{tab:action_rules} and \ref{tab:example_norms}).

Importantly, even with a good norm in place, defection is still an equilibrium.
Thus, it can be argued that reputation systems transform the problem of cooperation into a simpler  stag-hunt-like problem \cite{peysakhovich:2018}, with good and bad equilibria -- leading to cooperation and defection respectively.
This literature assumes that the set of strategies are predefined, and explored by agents in a completely random fashion. 
We are interested in relaxing this assumption by using RL.

\section{Learning to Use Reputations \label{easyproblem}}

\subsection{Q-Learning for Learning Reputation Mechanisms}
 
RL algorithms learn a policy from repeated interactions with the environment and attempt to balance exploration and exploitation to maximize rewards. Unlike in EGT, RL agents do not choose from a fixed set of existing strategies, but learn instead to take actions given environment observations.
% \item what are actions
% \item learning actions: tabular, states, actions...
Here, we train our agents with tabular $Q$-learning \cite{watkins:1992}. The policy, $\pi_i$ of agent $i$ is represented by a table of state-action values  $Q_i(s, a)$. 
While learning how to react to the reputations of others, states correspond to opponents' reputations. Actions are naturally cooperate or defect.

A single episode consists of a sufficiently large number of rounds $K$, where agents are randomly match to play a PD game using their reputations.
The policy of agent $i$ is an $\epsilon$-greedy policy and is defined by 
\begin{equation}
    \pi_i(s) = 
    \begin{cases}
        \argmax_{a \in \mathcal{A}} Q_i(s, a) & \text{with probability } 1-\epsilon \\
        \mathcal{U}(\mathcal{A}) & \text{with probability } \epsilon\\
    \end{cases}
\end{equation}

\noindent where $\mathcal{U}(\mathcal{A})$ denotes a uniform distribution over actions. Agents collect a set of trajectories $\{(s, a, r, s')_k : k = 1,...,K\}$ by interacting with the environment and store them in a memory buffer $\mathcal{M}_i$. When learning, it updates its policy using these experiences according to 

\begin{equation}
    Q_i(s,a) \leftarrow Q_i(s, a) + \beta [r + \gamma \max_{a' \in \mathcal{A}_i} Q_i(s', a') - Q_i(s, a)]
\end{equation}

\noindent where $s$ is the current state, $a$ is the current action, $r$ is the reward and $s'$ is the next state. As agents learn over time the environment appears non-stationary from any agent's perspective. 
To account for other agents learning, the information stored in each agent's memory buffer is refreshed every episode to ensure it remains relevant to the current transition dynamics of the environment.

In this environment, agents are both matched in rounds against other agents and must choose to cooperate or defect but must also judge the interactions of others and label agents as ``good'' or ``bad''. 
Agents do not accrue rewards for passing judgment and must coordinate how they assign reputation to others purely from the rewards received when cooperating or defecting with other agents. It might be difficult for agents to learn independently as they must coordinate on how to interpret reputations as well as how to assign them to others based on their behavior while simultaneously learning to cooperate or defect. 

We consider two scenarios. In our first scenario, agents learn how to react to the reputations of others, with an effective social norm being enforced by a central party -- this is discussed next. By analyzing the Q-values learned by the agents we can represent each agent's policy, $\pi_i$, as an equivalent strategy using the bitwise interpretation in Table \ref{tab:action_rules}. In the second scenario we consider mechanisms for learning from individual experience.

\begin{figure}[t]
\center
  \includegraphics[scale=0.55]{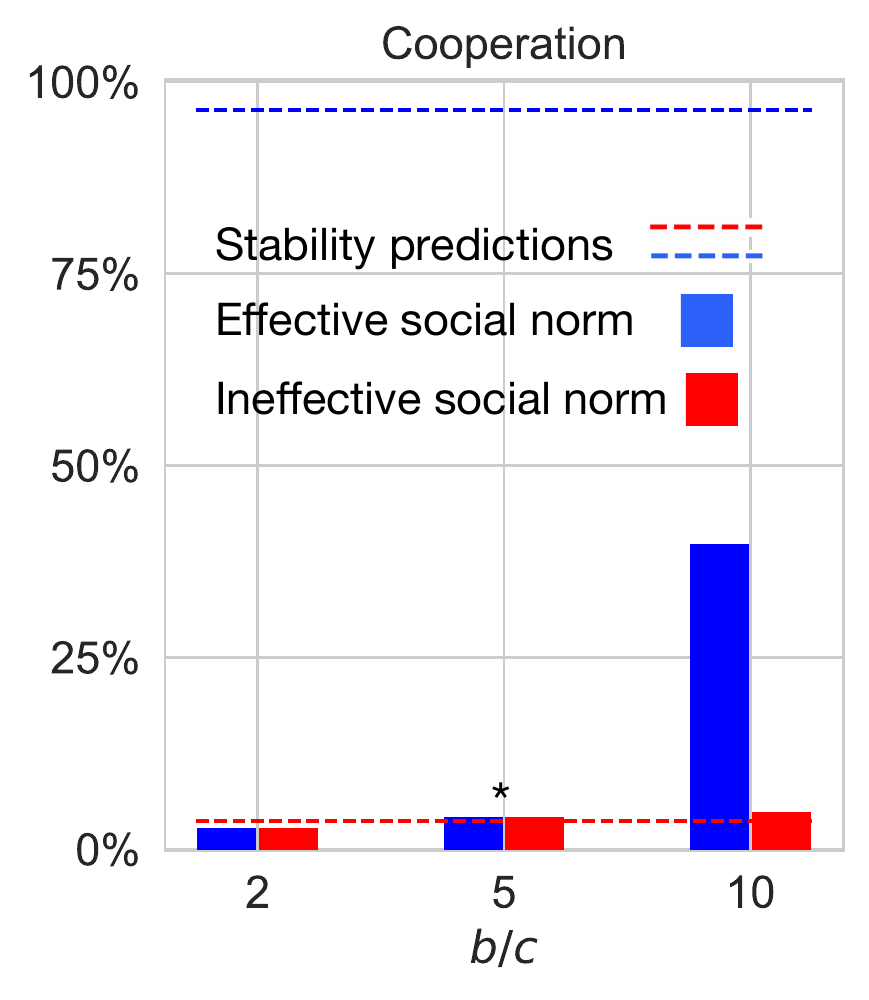}
  \caption{Standard Q-learning achieves substantially less cooperation that what is predicted with stability analysis.}
  \label{fig:stability_predictions}
  
\Description{This plot is a bar-chart that shows the amount of cooperation that is achieved using Q-learning compared to what is predicted with stability analysis in the presence of either an effective of ineffective social norm. We test this for various benefit-to-cost ratios which is determined by the rewards in the payoff table. When the benefit-to-cost ratio is low (2 and 5), Q-learning does not achieve cooperation which is similarly predicted with stability analysis. However, when benefit-to-cost ratio is 10, stability analysis predicts close to 100\% of outcomes to be cooperative when using an effective social norm, but in reality it is closer to 40\%. X-axis of this graph represents b/c ratio with values 2,5 and 10. Y-axis represents percentage of outcomes that end in cooperation from 0 to 100\%. Scores with an effective social norm are represented by a blue bar while scores with an ineffective social norm are represented by a red bar.}
\end{figure}

\subsection{Convergence to Inefficient Equilibria}

%\begin{itemize}
%	\item Gist is it doesn't work
%\end{itemize}

This effective norm (norm $9$), combined with an action rule that reacts to reputation (e.g., action rule $5$), makes cooperation stable (see Section \ref{stability}). 
Stability predictions expect this norm to maintain cooperation and, with social learning and stochasticity, a system with about $10$ agents will reach as much as $75\%$ cooperation for a benefit to cost ratio of $5$ \cite{santos:2016}. 

Can agents using RL learn to play the strategies that combine with stern judging to maintain cooperation?
We fix norm $9$, in a centralized system and allow agents to adjust their policies following the algorithm described above.
We start with a population of $10$ agents. 
The reputation dynamics simulation is ran for $10.000$ episodes, each one comprising $K = 200$ random encounters in the population. 
We set the reputation assignment error to $\chi = 1 \times 10^{-3}$ and, fixing the cost of cooperation to $c=1$, vary the benefit.
The learning rate is set to $\beta = 1 \times 10^{-2}$, $\gamma = 0.99$ and $\epsilon = 10 ^{-1}$. $\epsilon$ is kept at a fixed value to account for the changing behavior of other agents.
Results are averaged over $20$ different random seeds for each parameter set.
We measure the average reward in the whole population during the last half of the episodes, taking the average, weighted by $b$, as the level of cooperation achieved.
The results are shown in Figure~\ref{fig:stability_predictions}.

Strikingly, RL agents fail to reliably achieve cooperation, even in the presence of an effective social norm. 
Only when cooperation is very cheap, at a benefit to cost ratio of $10$, a small $40\%$ cooperation is maintained.
At a benefit to cost ratio of $5$, the differences between an effective social norm, and an ineffective social norm are negligible.
This is in stark contrast to what is expected from stability analysis \cite{ohtsuki:2006}, or even stochastic predictions relying on social learning.
For a system with $10$ agents, at a benefit to cost ratio of $5$, a model based on EGT techniques predicts as much as $70\%$ cooperation based on the same social norm.

This result can be explained by the fact that defection is still an equilibrium, even with an effective social norm.
Agents in this setup reliably fail to use the reputation information, converging on a purely defecting strategy that ignores reputation.
Effectively, the reputation system transforms a difficult prisoner's dilemma into stag-hunt like game with efficient (cooperation) and inefficient equilibria.
It has been reported before that in these situations, RL algorithms can fail to converge to desirable equilibria in the absence of intrinsic rewards or changes to the environment \cite{peysakhovich:2018}.

This outcome also reflects what typically happens when RL agents are trained to play with one another in a social dilemma: (i) learning does not account for potential changes in other agents' strategies and so defection is seen as more valuable if the environment were to remain unchanged, and (ii) agent behaviors are initially volatile and so observations are not fully representative of agent strategies.
This remains the case, even with a coordination system such as that introduced by the reputation mechanism, and even with a simple game with binary reputations.

More generally, the fact that agents trained with Q-learning cannot learn cooperation even with an effective social norm indicates a gap between social and individual learning. We propose two solutions to this problem to improve agent coordination on the meanings of reputation labels and encourage cooperation through introspective rewards.
These are discussed next.

\section{Steering agents towards efficient equilibria \label{fixes}}

We now set to design a mechanism to steer agents towards the efficient equilibrium, while retaining the main feature of learning from individual experience.
To do this, we take cues from the cooperation literature in EGT \cite{nowak:2006a}, where cooperation is enabled by  correlated interactions, allowing cooperative types to meet each other more frequently.
This guarantees that the benefits of cooperation are disproportionately shared only among those cooperating.
We propose two implementable ways to achieve this in the case of reputations.
These ideas are discussed in Sections~\ref{seeding}and~\ref{introspection}, respectively. 
In the remaining experiments we analyse the outcomes using a challenging value $b/c = 5$.

\subsection{Seeding Agents to Improve Coordination \label{seeding}}

Seeding agents with reciprocal strategies has the potential to increase the reward to autonomous cooperative ones.
This approach has been studied before in  norm-emergence scenarios \cite{borglund:2018, sen:2007}. In particular, \cite{sen:2007} considers a simpler case where norms are equivalent to strategies and only two norms are available. In our case, norms are coordination devices through reputation. This setup is richer because more than two norms are available --- clearly making the problem more difficult.

% This is akin to changing the distribution of opponents in a way that allows reciprocators to be more often that warranted by the agents's distributions on the receiving end of cooperation.

 To alleviate the burden of coordination, we look to encourage agents to coordinate around a specific equilibria by introducing fixed agents into the population that play action rule $5$ (i.e., $0101_2$): I cooperate only when my co-player has a good reputation. 
 Combined with social norm $9$, an agent that plays action rule $5$ is guaranteed to always have a good reputation. By specifically rewarding agents with good reputations and penalizing agents with bad reputations, it encourages agents to play strategies that, in turn, give them good reputations, thereby facilitating coordination. %{\color{red} The concept of utilizing fixed agents was also introduced in \cite{sen2007emergence}  and had success in steering the learners of the society towards a particular norm, however, the setup was significantly simpler with only 2 possible norms in a one action coordination game. Moreover, we investigate these mechanics in a game where agents must learn to coordinate their reputation assignments across 16 possible norms while simultaneously learning strategies to play a Dilemma game.} 

\begin{figure*}[hbt]
\center
  \includegraphics[scale=0.7]{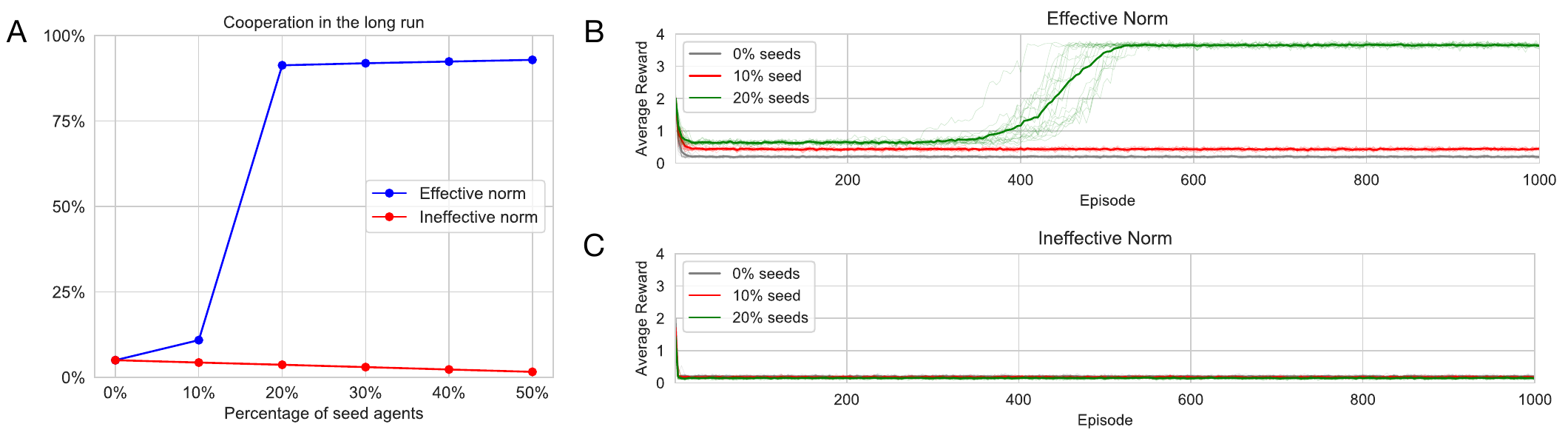}
  \caption{Seeding agents to promote cooperation. Panel A shows the average cooperation in the last half of the episodes, as measured by the proportion achieved of the maximum cooperation payoff. Panel B shows typical learning trajectories for the population of agents using an efficient social norm, highlighting the average trajectory in bold. Panel C shows the results for norm $0$.}
  \label{fig:seeding}
    \Description{3 panels are displayed in this figure. 
  
  Panel A: This line plot shows cooperation in the long run. The x-axis represents the percentage of seeded agents and ranges from 0\% to 50\% seeded agents. The y-axis represents the percentage of cooperation from 0\% to 100\%. A line is used to track the amount of cooperation against the percentage of seeded agents. When there is an effective norm a blue line is used and when there is an ineffective norm a red line is used. The blue line starts at 10\% and reaches almost 90\%. The red line starts at 10\% and approaches 0\%.
  Panel B: This line plot shows the amount of reward achieved over 1000 episodes for different amounts of seeded agents in the presence of an effective norm. The x-axis represents the number of episodes of training from 0 to 1000. The y-axis represents the average reward of agents. A grey line represents the amount of reward achieved at each episode when 0\% of the agents are seeded. This starts at 2.5 and quickly drops to 0. A red line represents when 10\% of the agents are seeded. This starts at 2.5 and quickly drops to 0. A green line represents when 20\% of the agents are seeded. This hovers near 0\% for 400 episodes and then gradually climbs up. After approximately 550 episodes the average reward achieved is approximately 3.75 and it stays there until the end of the simulation.
  Panel C: This is the same as Panel B except for an ineffective norm instead of an effective norm. All the lines follow the same trajectory starting at 2.5 and quickly dropping to 0. They remain at 0 average reward until the end of the simulation.
  
  }
\end{figure*}

We now vary the number of seeds $k$, i.e.,  we fix  $k$ agents  using action rule $5$ (i.e., $0101_{2}$), in an environment where $N-k$ agents are learning.
 For $k=N$ we recover a simple single-agent learning problem. 
 We will show that a small proportion of seeds $k$ is enough to successfully promote cooperation.
We run experiments with $N = 10$ agents that learn for $50.000$ episodes.  We focus on the challenging problem where $b/c=5$, and vary the number of seed agents $k$, expressed as proportion of the whole population.
To benchmark the effects of seeding agents, we consider the effective social norm ($9$), as well as norm $0$, which completely disregards the value of reputations.
Other RL parameters are kept as above.
The results are summarized in Figure~\ref{fig:seeding}.

Figure 3A shows how $k/N$ affects cooperation in the long run. It measures the average cooperation in the last half of the episodes, as measured by the proportion achieved of the maximum cooperation payoff.
There is a sudden shift, when $20\%$ of the agents are fixed reciprocators, steering the population towards cooperation reliably.
Typical learning trajectories are shown in Figures 3B and 3C, for the effective social norm, and the norm that disregards reputations respectively.
The results for a norm that ignores reputation shows that the seeding helps coordination when an effective norm is in place.
We note that the role of fixed reciprocators is to both regulate the amount of defection that learning agents can ``get away with'', while also also stabilizing the learning process by reducing the variance in the outcomes. 
We need about 20\% of the population to be reciprocators in order to converge to cooperative outcomes when the other 80\% of the agents are learning using Q-learning -- this result holds for different population sizes, at the same proportion of fixed agents. 
% Slightly confused about this sentence.
This is intuitive since the threshold that biases the random matching to guarantee reciprocation needs to stay constant when the system size is increased.

\begin{figure}[hbt]
\center
  \includegraphics[scale=0.73]{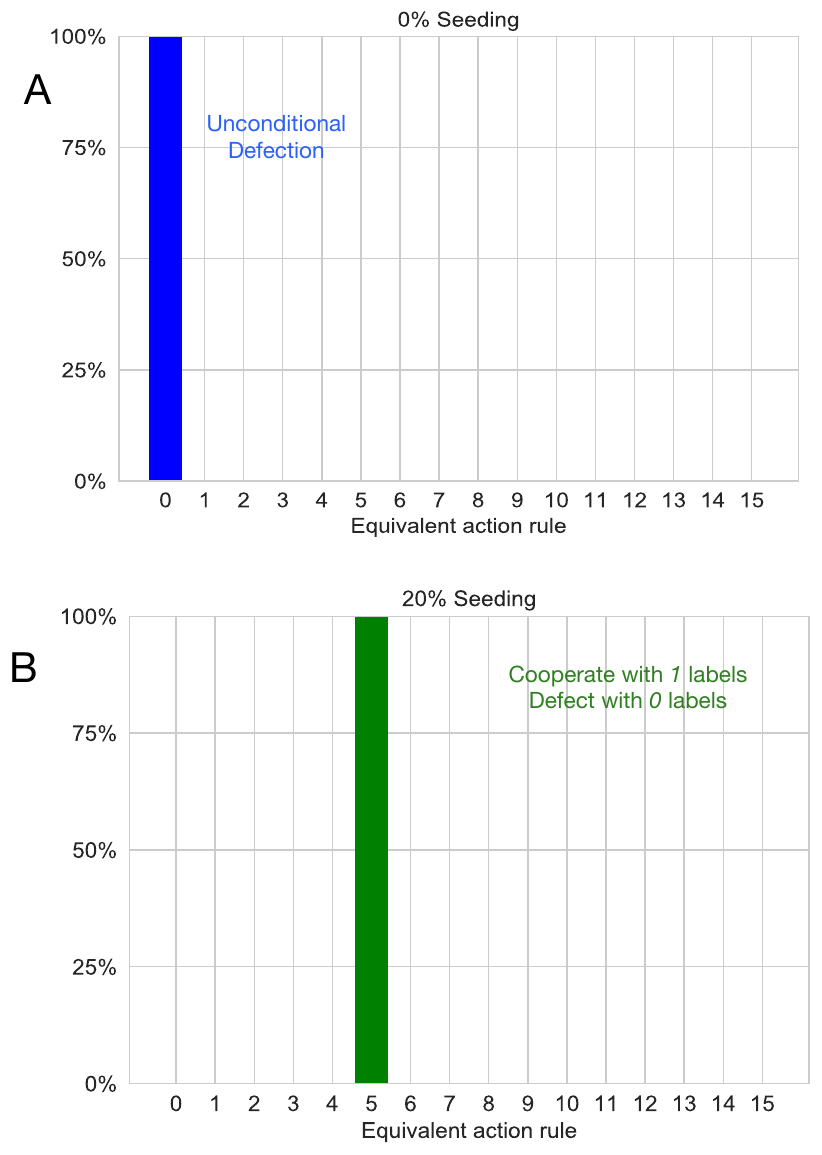}
  \caption{\label{strategies_seeding} Learned policies with seeding -- counting the frequency of strategies equivalent to the policies the algorithm has converged to.}
  
  \Description{2 panels are displayed in this figure:
	Panel A: This is a bar chart that shows which action rule is learned when 0\% of the agents are seeded. On the x-axis are the 16 possible action rules that can be learned. On the y-axis is the percentage of agents that learn this action from 0 to 100\%. There is one bar, in blue, that shows that 100\% of agents learn action rule 0 which is unconditional defection.  
	
	Panel B: This is the same as Panel A except for when 20\% of the agents are seeded. There is also one bar, in green, that shows that 100\% of agents learn action rule 5 which is to cooperate with agents who have label 1 and defect with agents who have label 0.
  }
\end{figure}

Many policies can represent strategies resulting in widespread cooperation. 
An obvious outcome is that agents just cooperate all the time regardless of what their opponents strategy is (action rule $15$). 
This would ultimately lead to maximizing the total reward if all agents stick to unconditional cooperation
This is crucially not the case when seeding reciprocators.
With an effective social norm, cooperating with agents who have bad reputations will beget a negative reputation. 
Thus, unconditional cooperators are not stable. 
Instead, agents learn to play in a way that warrants them a positive reputation, in turn maximizing the reward, i.e, using a reciprocal strategy.
Figure \ref{strategies_seeding} shows how in the long term, in the presence of fixed agents, RL agents converge to a policy equivalent to the reciprocal strategy that matches the fixed agents being seeded.
Without seeding, unconditional defection is the prevalent outcome.

In summary, assuming an exogenous effective social norm, seeding reciprocal agents (with action rule $5$) helps those learning to coordinate on the ``good'' label.
Evidence of this is that the mirror action rule $10$, that coordinates by limiting cooperation to those with label $0$, is not present in the long run.
While $20\%$ may not be a high demand in the proportion of seeded agents in certain circumstances, there may be scenarios where the level of decentralization does not allow for the system to have a predetermined number of fixed good agents e.g., when they are required to be physically present. However, there are many scenarios where it may be cheap given their impact such as instantiating software bots in a virtual environment.
We next discuss an alternative solution based on intrinsic rewards.

\subsection{Introspective Rewards \label{introspection}}

The idea of intrinsic rewards incorporates psychological insights from motivation into learning, by considering not only the external rewards provided by the environment, but also rewards that are intrinsic to agents \cite{barto:2013}.
This idea has been used before in cooperation problems with reinforcement learning, by endowing agents with a taste for curiosity \cite{jaques:2019} or with other-regarding preferences \cite{hughes:2018}.

Here, we use a simple principle for intrinsic rewards.
Agents care about what their policy would do to an agent like themselves.
Therefore, we consider this as a form of \textit{introspection}.
Thus, the extrinsic reward and the value of introspection are weighted with a linear combination with parameter $\alpha$.
An agent's $i$ reward is then
$$R_i = \alpha U_i + (1-\alpha) S_i$$
where, $U_i$ is their payoff in a particular encounter, and $S_i$ refers to the payoff they would get facing themselves.
The intuition for this is that agents would prime policies that would be effective when playing against agents like themselves.
The parameter $\alpha$, in $[0, 1]$, is used to represent the level of introspection.

While the self-encounter leading to the intrinsic reward $S_i$ still uses  the agent's reputation as an input, the actions the agents taken in during self-play do not affect their reputations. 
This ``simulated self-encounter'' does not contribute to the state of the game, but it is only used to generate the intrinsic reward.

Aside from the intuition of introspection, this mechanism also has a justification previously explored in the EGT literature. 
Parameter $\alpha$ can also be conceived as regulating the matching, and priming interactions among alike types; i.e., assortative matching \cite{bergstrom:2003a}. 
If meetings between agents are ``random'', those defecting will on average get higher rewards than cooperators; but when matching is assortative,  cooperators are more likely to meet cooperators than defectors -- therefore, the cost of cooperation is repaid by a higher probability of playing against a cooperating opponent \cite{vanveelen:2012}.

\begin{figure*}[hbt]
\center
  \includegraphics[scale=0.70]{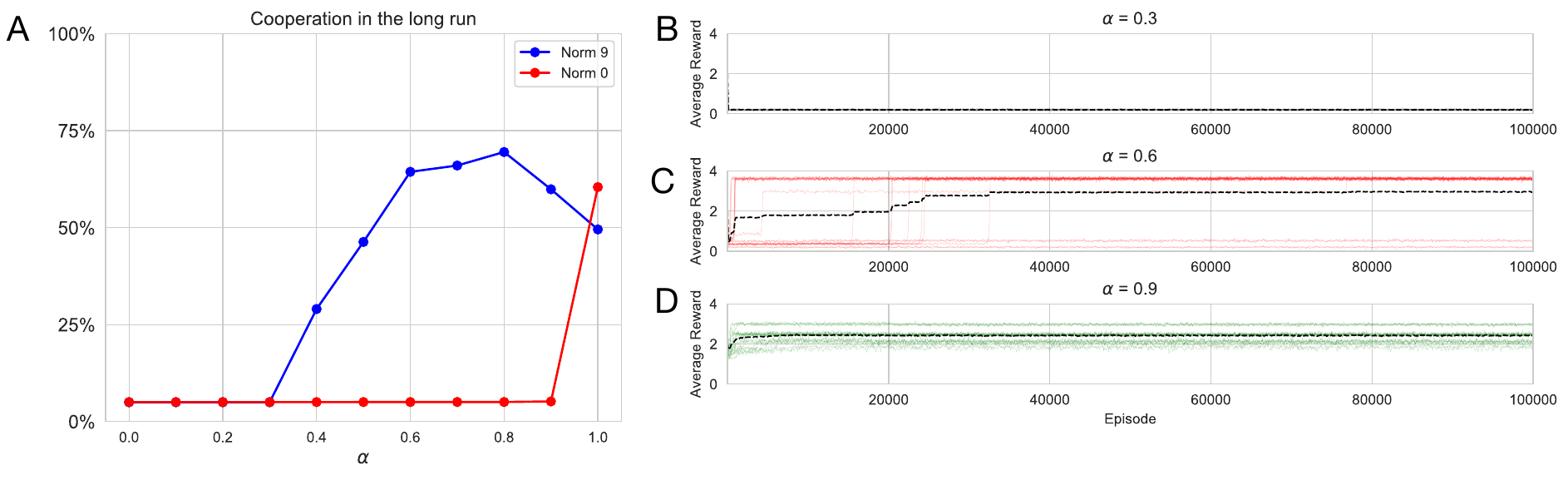}
  \caption{Seeding agents to promote cooperation \label{fig:introspective}. Panel A shows the average cooperation in the last half of the episodes, as measured by the proportion achieved of the maximum cooperation payoff as a function of the level of introspection. Panels B, C and D show typical learning trajectories for the population of agents using an efficient social norm with different levels of introspection $\alpha$. We  highlight the average trajectory as a dotted line. }
  
  \Description{This figure contains 4 plots:
  
  Panel A: This line plot shows cooperation in the long run. The x-axis represents a range of $\alpha$ values from 0 to 1.0, i.e. the amount of introspection that agents do. The y-axis represents the percentage of cooperation from 0 to 100\%. A line is used to track the amount of cooperation against the value of $\alpha$. When there is an effective norm a blue line is used and when there is an ineffective norm a red line is used. The blue line starts at 10\% and reaches almost 90\%. The red line starts at 10\% and approaches 0\%.
  
  Panel B: This line plot shows the amount of reward achieved over 10,000 episodes for $\alpha$ = 0.3. The x-axis represents the number of episodes of training from 0 to 10,000. The y-axis represents the average reward of agents. A grey line represents the amount of reward achieved at each episode when 0\% of the agents are seeded. This starts at 2.5 and quickly drops to 0. 
  
  Panel C: This is similar to Panel B except for $\alpha = 0.6$.
  
  Panel D: This is similar to Panel B except for $\alpha= 0.9$.
  }
\end{figure*}

The results are summarized in Figure~\ref{fig:introspective}. 
Parameters are as discussed above, but noting a larger variance in the rewards distribution we allow the learning to run for a larger number of episodes.
Panel A shows how different levels of introspection affect cooperation in the long run, for an effective and ineffective social norm. 
For $b=5$, the benefits of the intrinsic reward kick in, raising the level of cooperation.
Crucially, this curve is close to $0.5$ at $\alpha=1$, since the maximum level of introspection completely drowns out the signal from the environment. 
Cooperation peaks at $\alpha =0.8$. We have a decline in cooperation for larger values, driven by the noise associated with overemphasizing the intrinsic feature of the reward.
Panels B, C and D show typical learning trajectories for agents in an environment with the effective social norm. 
The average trajectory is shown as a dotted line.

\begin{figure}[t]
\center
  \includegraphics[scale=0.73]{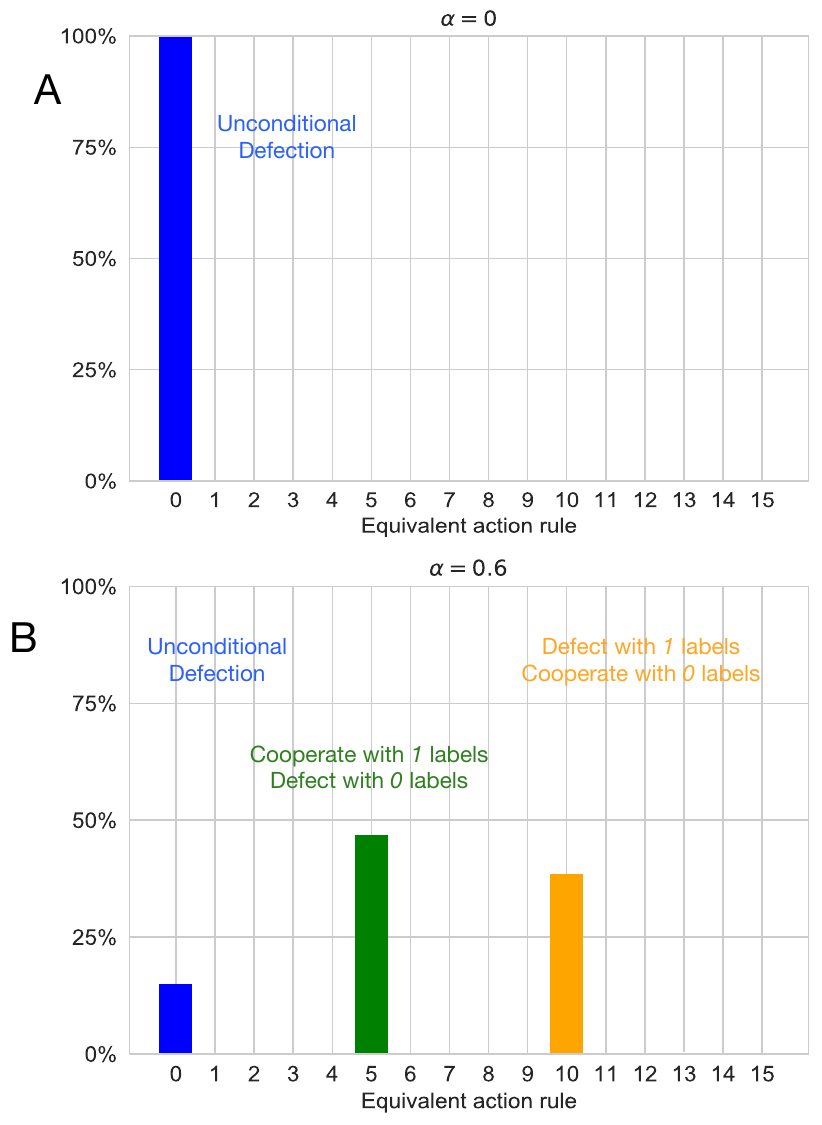}
  \caption{
\label{strategies_alpha} Learned policies with introspective rewards -- counting the frequency of strategies equivalent to the policies the algorithm has converged to.}
  \Description{2 panels are displayed in this figure:
	Panel A: This is a bar chart that shows which action rule is learned when $\alpha = 0$. On the x-axis are the 16 possible action rules that can be learned. On the y-axis is the percentage of agents that learn this action from 0 to 100\%. There is one bar, in blue, that shows that 100\% of agents learn action rule 0 which is unconditional defection.  
	
	Panel B: This is the same as Panel A except with $\alpha=0.6$. There are 3 bars on the chart to show how many agents learn each action rule. 12.5\% of agents learn action rule 0 represented with a blue bar. 45\% of agents learn action rule 5 represented with a green bar. 37.5\% of agents learn action rule 10 represented with a yellow bar. Action rule 10 is to defect against agents with label 1 and cooperate with agents with label 0 (reverse of action rule 5).
  }
\end{figure}

As reputation mechanisms turn a PD with a single non-efficient equilibrium (i.e., defection), into a game where potentially many equilibria arise depending on how individuals use the reputations for coordination.
In particular, agents need to coordinate on reacting to reputation signals.
While the introspective reward encourages cooperation, it does not help the agents to solve the signal coordination problem.
Action rule $5$, prominent in the results of Section \ref{seeding}, has a mirror action rule whereby agents defect with those labeled ``good'' ($1$), and cooperate with those labeled ``bad'' ($0$). 
When all agents decide on a label, both action rules can engender cooperation when combined with norm $9$. 
Figure \ref{strategies_alpha} shows that agents are sometimes divided on these two action rules, failing to cooperate consistently, and occasionally opening the door for unconditional defection.
While conditional strategies are used almost $90\%$ of the time for the optimal introspection level ($alpha=0.6$), cooperation can only go as high as $75\%$ with introspective rewards.
Combining seeding and intrinsic rewards can have a synergistic effect, with only $5\%$ seeding and $\alpha=0.6$ required to guarantee at least $90\%$ cooperation for $b/c=5$. %This paragraph is offered without graphical evidence.

We have shown how intrinsic rewards can increase cooperation, but are not enough to resolve the coordination problem behind reputation labels.
All of these results also assume that a central enforcer administers the judgement of reputation.
In other words, the agents are assumed to stick to a social norm that is effective.
This has limitations, because it requires mechanisms to solve the corresponding coordination problem of reputation assignment, or assumes levels of centralization where social norms can be enforced.
We next discuss a fully decentralized scenario, where a central enforcer is not required.

\section{Learning to assign reputations \label{hardproblem}}

The problem of reputation becomes harder when social norms are no longer centralized \cite{xu:2019}.
Instead, it is assumed that each agent can judge the reputations of other after each encounter.
So for every encounter, a third party is chosen -- randomly from the population -- to judge the reputations of the two parties involved in each interaction.
This reflects a completely decentralized system where enforcement of the social norm is not possible.
Thus, the next step is learning to coordinate not just how to react to others' reputations, but also how to assign reputations to other agents. 
Maintaining cooperation in this case requires agents to learn to assign reputations in a meaningful way, preserving information about other agents' strategies in the reputation labels. 

%NICOLAS: DESCRIBE TABULAR LEARNING IN THIS CASE. Two tables, what's the space for reputations... interestingly, rewards for reputations are not immediate... is there any similar problems?
Once again, agents do not choose from $16$ available norms to judge encounters between agents, but learn to assign a reputation $0$ or $1$ based on the actions agents take and the reputation of their co-player. The dimensionality of the Q-tables increases to accommodate for the new states and actions but the learning rate is kept fixed at $\beta=1e-2$ with $\epsilon=0.1$. Agents do not accrue rewards for judging the interactions of others and must coordinate how they assign reputation purely from the rewards received when cooperating or defecting with other agents. 
This is hard because agents must rely on others to assign informative reputations to agents and do not directly receive rewards for doing so.

%This is hard because agents do not directly receive credit for assigning an informative reputation to an agent since they are not guaranteed to interact with that agent themselves after but are instead randomly paired with any agent in the population. 

\begin{figure}[hbt]
\center
  \includegraphics[scale=0.245]{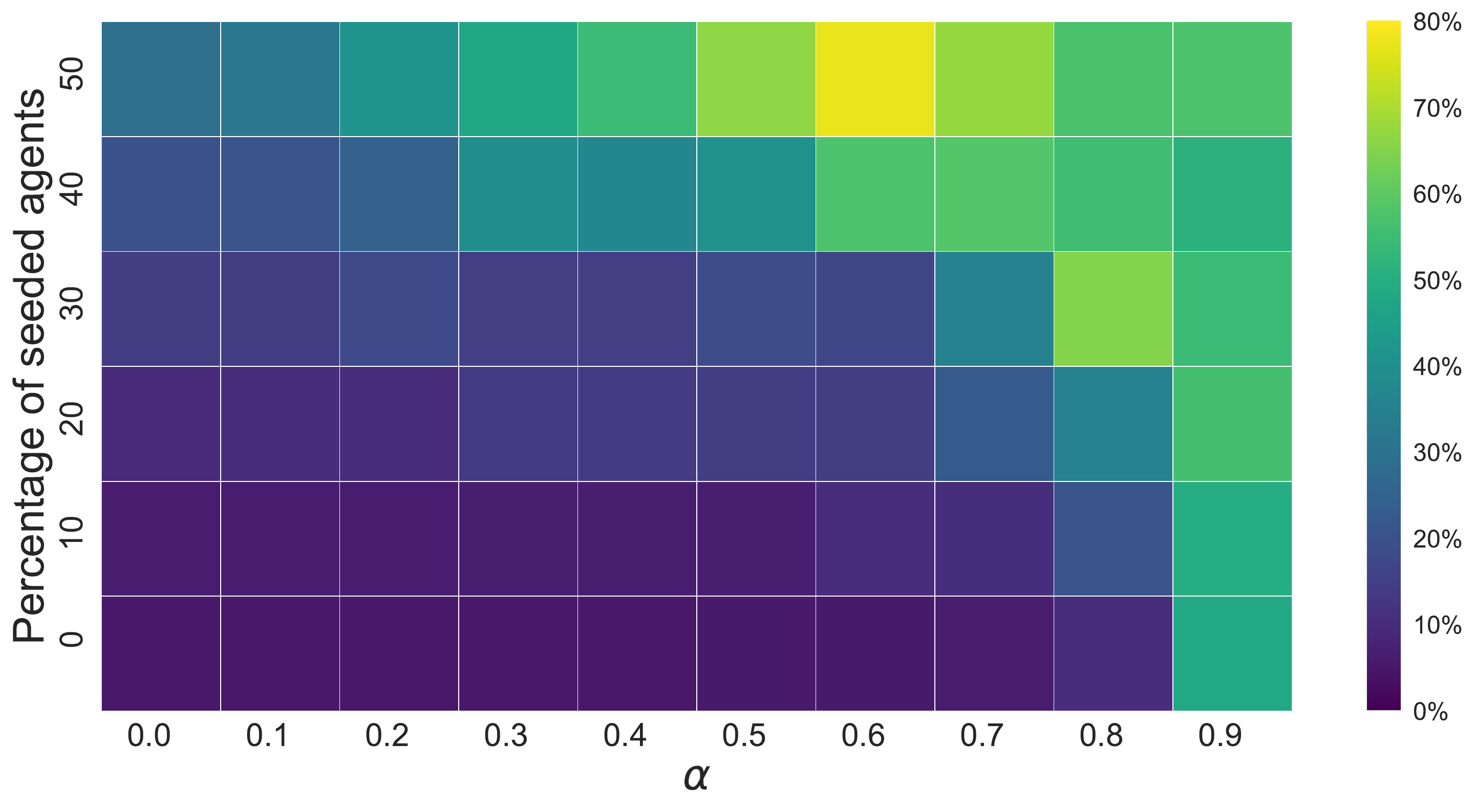}
  \caption{Introspective reward and seeded agents recover cooperation for fully decentralized reputations. \label{decentralised_cooperation_matrix}}
  \label{cm_payoffs}
  
  \Description{This is a confusion matrix to show the amount of cooperation recovered when there are X seeded agents and and introspection value of $\alpha$. On the y-axis we display the percentage of agents from 0 to 50\%. On the x-axis we display $\alpha$ ranging from 0 to 0.9. With 50\% seeded agents and $\alpha$ of 0.6 we recover 80\% cooperation which is the highest amount of cooperation that can be recovered from these values. }
\end{figure}

Figure~\ref{decentralised_cooperation_matrix} shows results for the decentralized problem.
Without introspection or seeded agents to improve agent coordination, agents do not learn to cooperate under these circumstances and converge to inefficient equilibria. 
We can immediately see the impact of an effective norm as agents quickly learn a defecting strategy that defects unconditionally. 
% \textcolor{red}{This renders reputation meaningless and learning an appropriate social norm becomes even more difficult.}
This further complicates the problem as it renders reputation meaningless and learning an appropriate social norm becomes difficult.

% Experimental results

As mentioned previously, seeded agents encourage coordination on the reputation signal while introspection encourages cooperation. 
Introducing our two mechanisms independently has middling success as neither result in cooperation consistently emerging. The added complexity of coordinating on how to assign reputation still results in defecting strategies being overall more rewarding for agents and even seeding $50\%$ of the agents only results in $30\%$ cooperation between RL agents. 
While increasing $\alpha$ appears to work significantly better than seeding agents when norms are not fixed, the strategies that agents learn are not coordinated and dominated by the noise in the introspective reward signal as $\alpha$ approaches $1$. 

\begin{figure}[hbt]
\center
  \includegraphics[scale=0.7]{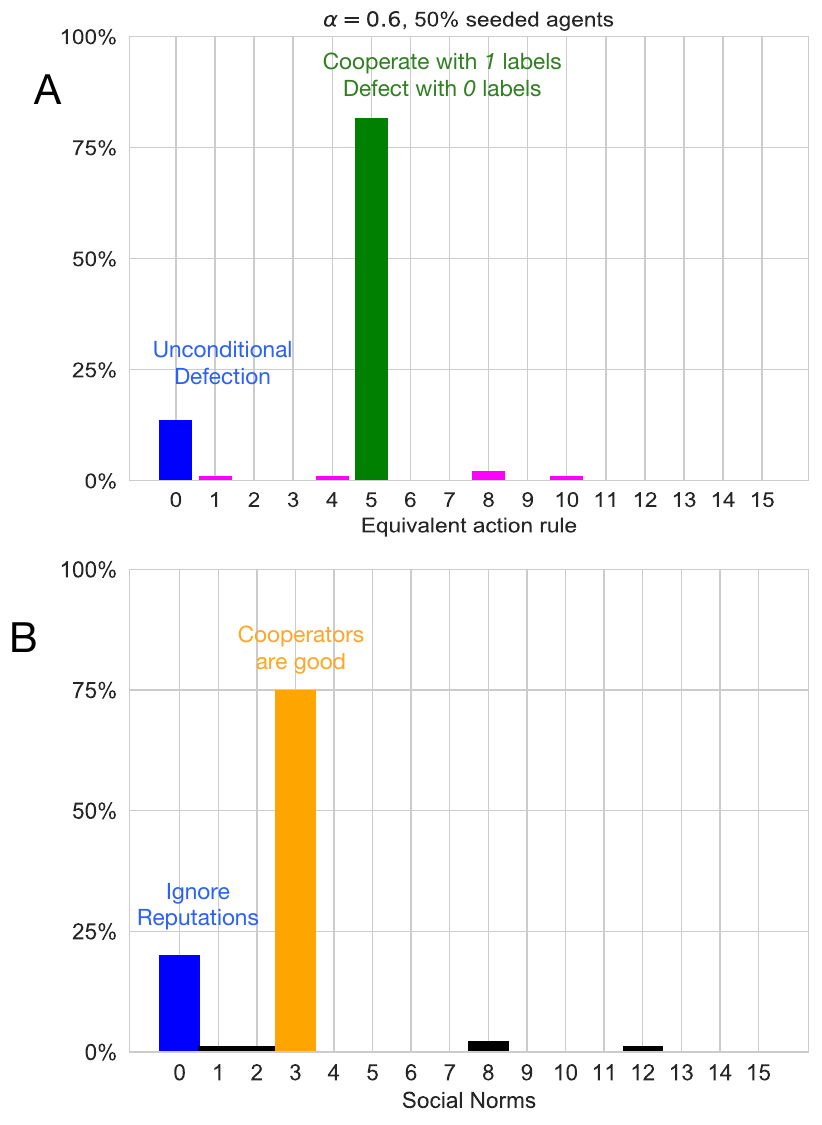}
  \caption{Introspective reward and seeded agents coordinate around social norm 3 and action rule 5 resulting in a cooperative, stable equilibrium.   \label{decentralized_reputation}}
  \Description{2 panels are displayed in this figure:
	Panel A: This is a bar chart that shows which action rule is learned when 50\% of the agents are seeded and $\alpha = 0.6$. On the x-axis are the 16 possible action rules that can be learned. On the y-axis is the percentage of agents that learn this action from 0 to 100\%. There is one bar, in blue, that shows that 12.5\% of agents learn action rule 0 which is unconditional defection. Another bar in green shows 80\% of agents learn action rule 5. In purple we display the bars with the remaining percentage spread out across the remaining action rules at very low percentages (sub 5\%)
	
	Panel B: This is the same as Panel A except on the x-axis are the 16 possible social norms that agents can learn. There are 3 bars on the chart to show how many agents learn each action rule. 20\% of agents learn to assign reputation according to social norm 0 represented with a blue bar. 75\% of agents learn action rule 3 represented with a yellow bar. In black we display the bars with the remaining percentage spread out across the remaining social norms at very low percentages (sub 5\%)
  }
\end{figure}

By combining these two mechanisms we can significantly improve the performance of the RL agents and have them coordinate around a social norm. Notably, with $50\%$ seeded agents and $\alpha=0.6$, agents can achieve $80\%$ cooperation. Although a large proportion of agents need to be seeded, we can see that this serves to curb agents from defaulting to defecting strategies enough that they can learn to assign meaningful reputations that are representative of agent strategies. This is shown in Fig \ref{decentralized_reputation}, panel B, where we can see that $75\%$ of the time, agents successfully coordinate around social norm 3. Unlike norm 9, norm 3 looks to identify agents according to their most recent action taken, labeling agents who have cooperated as ``good'' and agents who have defected as ``bad'' (see Table \ref{tab:example_norms}). In conjunction with norm 3, agents also learn action rule 5: cooperating with good agents and defecting against bad agents. Similarly to when norm 9 is used, the combination of norm 3 and action rule 5 results in a stable equilibrium that rewards cooperation while defectors are identified and punished. Unlike norm 9, agents who cooperate with defectors are not seen as bad agents and are not held responsible for their opponent's reputation. Instead an agent's reputation is determined exclusively by its own actions, independent of the opponent's reputation and is sufficient to represent each agent's behavior and guide RL agents towards a positive equilibria.  
The right combination of seeding and introspective rewards can recover up to $80\%$ cooperation in the fully decentralized case.
Full coordination remains a challenge.

\section{Conclusions} 
\label{discussion}

%General Conclusion
Reputation dynamics create difficult coordination and cooperation problems for independent learners. 
Agents trained using reinforcement learning fail to converge to efficient equilibria, even if an effective social norm is imposed with a reputation system. 
Just like in existing reputation systems with human actors, artificial agents have problems coordinating the effective use reputations. 

We have proposed two solutions to this problem.
%Seeding
Our first solution is to seed fixed agents whose task is to steer others towards coordination on the meaning of reputation labels.
Specifically, a mass of reciprocal fixed agents effectively helps RL agents to coordinate on a single label.
In turn, encouraging the conditional actions that foster cooperation while protecting it from defectors.

%Introspection
Introspection -- via intrinsic rewards -- entices agents to be more cooperative. 
This mechanism is theoretically grounded by models of assortative matching in EGT. 
The optimal balance of introspection and external rewards can recover a great deal of cooperation, but does not protect agents from not coordinating on reputation labels. 
%Combined result

These mechanisms show great potential for synergistic interactions. When combined, they successfully recover substantial levels of cooperation in fully decentralized scenarios.
%EGT conclusion
Our results also show that stability analysis and stochastic models of social learning, common in EGT, tend to over-estimate how much cooperation you can expect from the presence of reputation mechanisms. 
EGT models, and more generally models of human behavior, can learn from RL methods. 
Specifically, RL grounds the exploration process whereby agents discover strategies.
This process is often assumed ex-ante in evolutionary games.
As shown here, this can have an effect in model predictions.
Our results further underscore the differences between social and individual learning that are notable in the EGT and RL communities.
This work has been based on a rather basic scenario.
Future work may further explore how these solutions apply to more complex cooperation scenarios, including those beyond binary reputations. 
Grid-like worlds that are popular benchmarks in the RL community may also be interesting testbeds for understanding reputation.

%AI conclusion
Intelligent artificial agents are expected to be able to navigate social interactions and recognise efficient outcomes where multiple parties can benefit.  
Although a standard RL algorithm fails to converge to desirable equilibria, this can be amended by introducing successful mechanisms, which has been extensively investigated in the EGT literature.
We believe that this paper demonstrates the largely unexplored potential of combining techniques and methodologies from the RL and EGT communities in order to investigate open problems around cooperation and reputation dynamics in artificial, human, and hybrid societies.

%\newpage

%%%%%%%%%%%%%%%%%%%%%%%%%%%%%%%%%%%%%%%%%%%%%%%%%%%%%%%%%%%%%%%%%%%%%%%%

%%% The acknowledgments section is defined using the "acks" environment
%%% (rather than an unnumbered section). The use of this environment 
%%% ensures the proper identification of the section in the article 
%%% metadata as well as the consistent spelling of the heading.

%\begin{acks}
%If you wish to include any acknowledgments in your paper (e.g., to 
%people or funding agencies), please do so using the `\texttt{acks}' 
%environment. Note that the text of your acknowledgments will be omitted
%if you compile your document with the `\texttt{anonymous}' option.
%\end{acks}

%%%%%%%%%%%%%%%%%%%%%%%%%%%%%%%%%%%%%%%%%%%%%%%%%%%%%%%%%%%%%%%%%%%%%%%%

%%% The next two lines define, first, the bibliography style to be 
%%% applied, and, second, the bibliography file to be used.
\balance

\bibliographystyle{ACM-Reference-Format} 
\bibliography{aamas21}

%%%%%%%%%%%%%%%%%%%%%%%%%%%%%%%%%%%%%%%%%%%%%%%%%%%%%%%%%%%%%%%%%%%%%%%%

\end{document}